\documentclass[12pt,showpacs,amsfonts,superscriptaddress]{revtex4}
\usepackage{amsmath}
\usepackage{latexsym}
\usepackage{float}
\usepackage{amssymb}
\usepackage{graphicx}
\usepackage{textcomp}
\usepackage{hyperref}
\usepackage[dvipsnames,usenames]{color}
\usepackage{graphicx}% Include figure files
\begin{document}
\def\be{\begin{equation}}
\def\ee{\end{equation}}
\def\bfi{\begin{figure}}
\def\efi{\end{figure}}
\def\bea{\begin{eqnarray}}
\def\eea{\end{eqnarray}}
\newcommand{\bra}{\langle}
\newcommand{\ket}{\rangle}
\newcommand{\vk}{\vec{k}}
\newcommand{\rphi}{\widehat{\varphi}}
\newcommand{\iphi}{\widetilde{\varphi}}
\newcommand{\reta}{\widehat{\eta}}
\newcommand{\ieta}{\widetilde{\eta}}
\newcommand{\vphi}{\boldsymbol{\varphi}}
\newcommand{\eRk}{\emph{R}_{\vec{k},\eta}}
\newcommand{\eIk}{\emph{I}_{\vec{k},\eta}}

\title{Heat exchanges in  a quenched ferromagnet}

\author{Federico Corberi}
\affiliation {Dipartimento di Fisica ``E.R. Caianiello'',
%and INFN, Gruppo Collegato di Salerno,
 and CNISM, Unit\`{a} di Salerno,
Universit\`a  di Salerno, via Ponte don Melillo, 84084 Fisciano
(SA), Italy.}
 \affiliation{INFN, Gruppo Collegato di Salerno, 84084 Fisciano
(SA), Italy. }

\author{Giuseppe Gonnella}
\affiliation {Dipartimento di Fisica, Universit\`a di Bari and INFN,
Sezione di Bari, via Amendola 173, 70126 Bari, Italy.}

\author{Antonio Piscitelli}
\affiliation {Dipartimento di Fisica, Universit\`a di Bari and INFN,
Sezione di Bari, via Amendola 173, 70126 Bari, Italy.}

\author{Marco Zannetti}
\affiliation {Dipartimento di Fisica ``E.R. Caianiello'', and CNISM,
Unit\`{a} di Salerno, Universit\`a  di Salerno, via Ponte don
Melillo, 84084 Fisciano (SA), Italy.}

\begin{abstract}

The off-equilibrium probability distribution of the heat  exchanged
by a ferromagnet  in a time interval after a quench below the
critical point is calculated analytically in the large-$N$ limit.
The distribution is characterized by a singular threshold $ Q_C<0$,
below which a macroscopic fraction of heat is released by the $k=0$
Fourier component of the order parameter. The mathematical structure
producing this phenomenon is the same responsible of the order
parameter condensation in the equilibrium low temperature phase. The
heat exchanged by the individual Fourier modes follows a non trivial
pattern, with the  unstable modes at small wave vectors warming up
the modes around a characteristic finite wave vector $k_M$. Two
internal temperatures, associated to the $k=0$ and $k=k_M$ modes,
rule the heat currents through a fluctuation relation similar to the
one for stationary systems in contact with two thermal reservoirs.


\end{abstract}

\maketitle

Finding the principles underlying the probability measures of
non-equilibrium fluctuations is one of the most challenging and far
reaching open questions in modern statistical physics. Large
deviation theory, as recognized in the last years, gives a general
theoretical  framework for describing probability distribution
functions (PDF) in non-equilibrium states \cite{touchette}. However,
in spite of recent important developments \cite{Jona}, explicit
calculations, especially for interacting systems,  are limited to
few  cases \cite{touchette}. A major advance for non-equilibrium
stationary states has been the recognition of a general symmetry of
the PDF of certain observables, described  by the so-called
fluctuation theorems \cite{FL1,FL2}. For example, for a system
 in contact with reservoirs at two  inverse
temperatures $\beta _1<\beta_2$ \cite{FL3,FL4}, the probability
distribution $P(Q)$ that the  heat $Q$ flows from the first to the
second heat bath in a large time interval
 is related to  $P(-Q)$ by
 \be
\ln \frac{P(Q)}{P(-Q)} = (\beta_2-\beta_1){ Q} .\label{fr} \ee

Fluctuation behavior in  non-stationary states is by far less
understood. In the thoroughly investigated field of aging systems,
such as quenched ferromagnets or binary mixtures, disordered
materials and glasses, heat PDF  have been considered only
numerically in some specific disordered models \cite{crisrit} and,
recently, in an experiment for a brownian particle \cite{peliti}
trapped in an aging bath \cite{ciliberto}. Understanding the
properties of heat fluxes in these systems is of great importance
also for what concerns the notion of an {\it effective} temperature
\cite{KC}, which is expected to regulate such fluxes similarly to
what the ordinary temperature does in equilibrium.

In this Letter we address the latter category of problems, by
studying the probability distribution  of heat exchanges in a
ferromagnetic model quenched from a disordered state  to a final
temperature below the critical point. We do this by an
 exact calculation carried out on the time dependent Ginzburg-Landau
model  with an $N$-component vector order parameter in the large-$N$
limit \cite{bray}. Specifically, we find the analytical form of the
probability distribution $P(Q,t,t_w)$, where $Q$ is the heat
exchanged during the time interval $[t_w,t]$ following the quench.
Most interesting is the existence of a singular threshold $Q_C$,
such that for $Q < Q_C$ the macroscopic amount of heat $Q - Q_C$ is
entirely released by the zero wave vector mode. This comes about
through the same mechanism responsible of the transition to the low
temperature phase in the equilibrium version of the model
\cite{berlkac,condvs}. Furthermore, we find that  $P(Q,t,t_w)$
asymptotically obeys
a fluctuation relation akin to Eq. (\ref{fr}), even though the
system is not in a stationary state. This can be interpreted as due
to the heat exchanged between the condensing $k=0$ mode, lowering
the system energy as an effect of the ordering process, and the
modes at some finite characteristic wave vector $k_M$. The two
inverse temperatures playing the role of $\beta_1,\beta_2$ in
 Eq. (\ref{fr}) arise as the
typical energy scales associated to these two kinds of
non-equilibrium modes, and reduce
to the bath temperature
when the system is in equilibrium.

We consider a system of volume $V$, described by the Ginzburg-Landau
Hamiltonian
\begin{equation}
H[\boldsymbol{\varphi}]=\int_{V}d^dx\left[\frac{1}{2}
(\nabla\boldsymbol{\varphi})^2+\frac{r}{2}\boldsymbol{\varphi}^2+
\frac{g}{4N}\left(\boldsymbol{\varphi}^2\right )^2\right]
\label{energy}
\end{equation}
where $r<0$, $g>0$, and $\boldsymbol {\varphi} =
(\varphi_1,..,\varphi_N)$, is the $N$-component order parameter
field. Dynamics is governed by the Langevin equation
\be
\partial\varphi_{\alpha}/\partial t=-\delta H[\boldsymbol{\varphi}]
/\delta \varphi_{\alpha}+\eta_{\alpha}, \ee where
$\varphi_{\alpha},\eta_{\alpha}$ stand for
$\varphi_{\alpha}(\boldsymbol{x},t),\eta_{\alpha}(\boldsymbol{x},t)$
and the latter one represents the Gaussian white noise generated by
the thermal bath  with averages
$<\eta_{\alpha}(\boldsymbol{x},t)> = 0$ and  $ <\eta_{\alpha}
(\boldsymbol{x},t) \eta_{\beta}(\boldsymbol {x'},t')> = 2 T
\delta_{\alpha\beta} \delta(t-t') \delta(\vec x - \vec x')$. The
leading order of all quantities of interest in the $1/N$-expansion
can be obtained by replacing the above Hamiltonian with the
time-dependent effective one
\begin{equation}
{\cal H}[\boldsymbol{\varphi}]=\frac{1}{2} \int_{V}d^dx\left[
(\nabla\boldsymbol{\varphi})^2+ (r+gS(t))\boldsymbol{\varphi}^2 \right]- (NVg/4)S^2(t)
\label{energy.1}
\end{equation}
where $S(t)=\langle \boldsymbol{\varphi}^2(\boldsymbol{x},t) \rangle
/N$ must be computed self-consistently and angular brackets stand
for the average over both initial condition and thermal noise.
Due to space homogeneity, the quantity $S$ only
depends on time.  The remarkable feature of the large-$N$ limit is
that the dynamics generated by ${\cal H}$, although retaining all
the relevant features of the phase-ordering process, becomes exactly
soluble \cite{coniglio,bray}. By Fourier transformation one obtains
a decoupled set of formally linear equations of motion
\begin{equation}
\frac{\partial}{\partial t}\varphi_{\alpha}(\boldsymbol{k},t)=-\omega(k,t)\varphi_{\alpha}(\boldsymbol{k},t)+
\eta_{\alpha}(\boldsymbol{k},t)
\label{ev_eq}
\end{equation}
where the stiffness of each mode is given by $\omega(k,t)
=k^2+r+gS(t)$ and
$<\eta_{\alpha}(\boldsymbol{k},t)>=0$,
$<\eta_{\alpha}(\boldsymbol{k},t)\eta_{\beta}(\boldsymbol{k}',t')>
=2T\delta_{\alpha\beta}
\delta(\boldsymbol{k}+\boldsymbol{k}')\delta(t-t') $. Integrating
Eq.~(\ref{ev_eq}) and taking averages, the various observables can
be obtained \cite{noi2002}. In particular, the two-times structure
factor $C(k,t,t_w)= (1/V)\langle
\varphi_\alpha(\boldsymbol{k},t)\varphi_\alpha(-\boldsymbol{k},t_w)
\rangle$ with $t_w \leq t$ will play a relevant role in the
following.

The probability to release the heat $Q$ per component
in the time interval $[t_w,t]$ is defined by
\be
P(Q;t_w,t)\hspace{-.1cm}=\hspace{-.1cm}\int \hspace{-.1cm}
d[\boldsymbol{\varphi}_t] d[\boldsymbol {\varphi}_w] {\cal
P}(\boldsymbol {\varphi}_t,\boldsymbol {\varphi}_w)
\delta (Q-
\frac{{\cal H}[\boldsymbol {\varphi}_t]}{N}
+\frac{{\cal H}[\boldsymbol {\varphi}_w]}{N}),
\label{prob}
\ee
where
 ${\cal P}(\boldsymbol {\varphi}_t,\boldsymbol
{\varphi}_w)= \prod _{k,\alpha}{\cal
P}_{k,\alpha}(\varphi_{\alpha}(\boldsymbol{k},t),\varphi_{\alpha}(\boldsymbol{k},t_w))$
is the joint probability of the two configurations $(\boldsymbol
{\varphi}_t,\boldsymbol {\varphi}_w)$ at the times $t$ and $t_w$.
For a Gaussian process
\be
{\cal
P}_{k,\alpha}(\varphi_{\alpha,t},\varphi_{\alpha,w})= {\cal N}^{-1}
\exp \left \{- \frac{C_{tt}C_{ww}}{2V(C_{tt}C_{ww}-C^2_{wt})}
\left[\frac{\varphi^2_{\alpha,t}}{C_{tt}}+\frac{\varphi^2_{\alpha,w}}{C_{ww}}
-\frac{2C_{wt}\varphi_{\alpha,t}\varphi_{\alpha,w}}{C_{tt}C_{ww}}\right]\right
\}
\ee
where the short notation $C_{wt}\equiv C(k,t,t_w)$
(and similarly for $C_{tt},C_{ww}$)
has been introduced and
${\cal N}$
%=[2\pi V(C_{tt} C_{ww}-C_{wt}^2)]^{1/2}$
is the normalization. Next, using the representation
$\delta(x)=\frac{1}{2\pi i}\int_{-i\infty}^{i\infty}e^{-zx}dz$ of
the Dirac $\delta$-function and carrying out the integration in
Eq.~(\ref{prob}), we find
\begin{equation}
P(Q;t,t_w)=\int_{z_0-i\infty}^{z_0+i\infty}\frac{dz}{2\pi i}\,e^{V
h({\cal Q},z; t,t_w)} \label{p_di_q_con_modo_zero}
\end{equation}
where ${\cal Q}=\frac{Q}{V}$ is the heat density and the real
quantity $z_0$   is chosen in such a way that the integral is well
defined \cite{berlkac}. For large $V$ discrete sums over wave
vectors can be replaced by integrals, yielding
\be h({\cal
Q};z;t,t_w)=-z[{\cal Q}+gU(t,t_w)]-(1/2)\int_{\Lambda}\frac{d^dk}{(2\pi)^d}
\ln[1-zq(k,t,t_w)-z^2b(k,t,t_w)] \label{eqh} \ee
where
\be
U(t,t_w)=[S^2(t)-S^2(t_w)]/4,
\ee
\be
b(k,t,t_w)=\omega(k,t)\omega(k,t_w)[C(k,t,t)C(k,t_w,t_w)-C^2(k,t,t_w)],
\ee
 \be
 q(k,t,t_w)=\omega(k,t) C(k,t,t)-\omega(k,t_w) C(k,t_w,t_w),
 \label{qk}
 \ee
and the symbol
 $\int_\Lambda$ denotes an integral with an ultraviolet
cut-off $\Lambda$ related to the lattice spacing.

\begin{figure}[h]
\vspace{0.35cm}
    \centering
   \rotatebox{0}{\resizebox{.45\textwidth}{!}{\includegraphics{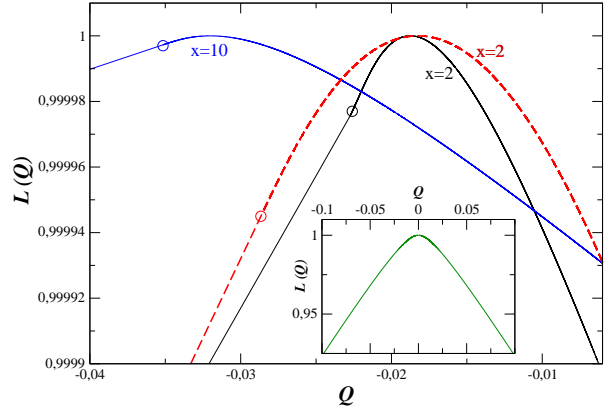}}}
   \caption{The large deviation function ${\cal L}({\cal Q};t,t_w)$ is plotted against
${\cal Q}$, in $d=3$, for $t_w=10$ and two values of $x$ for a
quench to  $T=0$ (continuous lines), or $T \approx  T_c/3$ (dashed
line). Circles represent ${\cal Q}_c$. In the inset the same
quantity is plotted in equilibrium at $T<T_c$.} \label{fig-2}
\end{figure}

Eqs. (\ref{p_di_q_con_modo_zero},\ref{eqh}) are completely general.
Different dynamical protocols are encoded into the correlation $C$.
We start our analysis from the simpler case in which the system is
in equilibrium at a generic  temperature $T$. In this case
$\omega(k,t)=\omega_{eq}(k)$  and $C(k,t,t)=C_{eq}(k)$ do not depend
on time due to stationarity, so from Eq. (\ref{qk}) $q= 0$ and
similarly $U= 0$. Moreover $C(k,t,t_w)= C_{eq}(k,t-t_w) =
(T/\omega_{eq}) \exp [-\omega_{eq} (t-t_w)]$ \cite{noi2002}, hence
$b= T^2[1-\exp[-2\omega_{eq}(t-t_w)]$. In the large $V$ limit, the
integral in (\ref{p_di_q_con_modo_zero}) can be computed by the
steepest descent method.
%{\color{red} Dove e' finita la dipendenza da t-tw?,
%ma poi l'integrale seguente diventa trivial?}
The saddle point equation ${\frac{dh}{dz}\vline}_{z=z^*}=0$ reads
\begin{equation}
{\cal Q}=\int_{\Lambda}
\frac{d^dk}{(2\pi)^d}\frac{z^*(t,t_w)b(k,t,t_w)}{1-{z^*(t,t_w)}^2b(k,t,t_w)}.
\label{saddle_point_eq}
\end{equation}
In order for $h$ in Eq. (\ref{eqh}) to be defined, it must be $z\neq
b(k,t,t_w)^{-1/2}$ $\forall k,t,t_w$. This, for $t-t_w \gg
1/\omega_{eq}(\Lambda)$, requires $-\beta < z_0 < \beta$,
with $\beta =1/T$. From the above expression of
$b$ one sees that the integral approaches infinity as $z^*\to \pm
\beta$
 so that Eq.(\ref{saddle_point_eq})  admits a
real solution $z^*$ for any value of ${\cal Q}$. The large deviation
function defined by
\be
P(Q;t,t_w)=\exp\{V{\cal L}({\cal Q};t,t_w)\}
\label{largedev}
\ee
is given by ${\cal L}({\cal Q};t,t_w)  = h({\cal Q};z^*;t - t_w)$
and is plotted in the inset of Fig. \ref{fig-2}. It is symmetric
and, for ${\cal Q}$ not too small behaves linearly, ${\cal L}\simeq
-\beta \vert{\cal Q}\vert$, since $z^*$ rapidly converges to $\beta$
as ${\cal Q}$ increases. Notice that the equilibrium temperature can
be read out from the singular points of $h$, which in turn regulate
the exponential decay of the tails of $P$.

Next, we consider the quench from infinite temperature to $T<T_c$,
starting with $T=0$. Since $q(k,t,t_w)$ will now play a central
role, let us comment on its physical  meaning. Using the normal
modes decomposition, the average energy at the time $t$ can be
written as $\langle {\cal H}  \rangle = \frac{N}{2}
\sum_{\boldsymbol{k}} \omega(k,t) C(k,t,t)-(NVg/4)S^2(t)$. This shows
that $q$ in Eq. (\ref{qk}) can be
interpreted as the average heat (per component) exchanged by the individual modes,
since the contributions due to $VgS^2/4$ become negligible at large
times as we will show below.

\begin{figure}[h]
\vspace{0.5cm}
    \centering
   \rotatebox{0}{\resizebox{.45\textwidth}{!}{\includegraphics{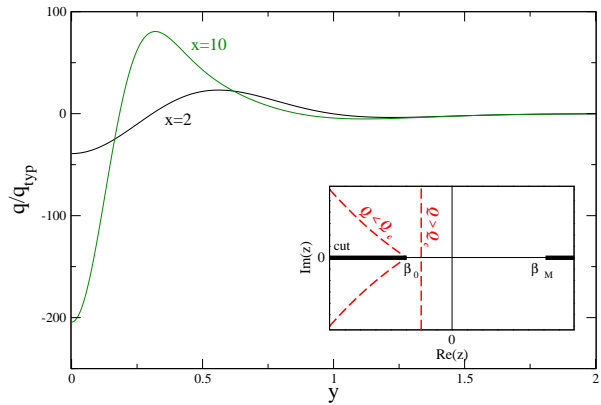}}}
   \caption{The quantity $q(k,t,t_w)$ is plotted in $d=3$ against $y$ (for large $t_w$)
and for two different choices of $x=t/t_w$. In the inset the
analyticity region of h (Eq.(\ref{eqh}))  and   the steepest descent
paths are plotted (dashed red).}
\vspace{-0.2cm}
\label{fig-1}
\end{figure}

In a zero temperature quench, one has
$C(k,t,t_w)=\sqrt{C(k,t,t)C(k,t_w,t_w)}$, which implies $b = 0$.
In what follows we will consider the large $t_w$ limit.
In this regimes one finds \cite{bray,coniglio,noi2002}
$S(t)=-r/g-d/(4gt)$ and
the dynamical scaling property ${L(t_w)^{-d}C(k,t,t_w)}=C(x,y)$,
where $x=t/t_w$, $y=kt_w^{1/2}$, $L(t_w)=(2t_w)^{1/2}$ is the
characteristic lengthscale at the age $t_w$ of the system,
and $C(x,y)=(4\pi )^{d/2}(-r/g)x^{d/4}\exp [-y^2(x+1)]$.
Hence,
using these results,
also $q$ can be written in the
scaling form
\be \frac{q(y,x)}{q_{typ}}=
(xy^2-\frac{d}{4})x^{\frac{d}{2}-1}e^{-2xy^2}-
(y^2-\frac{d}{4})e^{-2y^2} \label{eqqsing}
\ee
where
$q_{typ}(t_w)=-[r(8\pi)^{\frac{d}{2}}t_w^{d/2-1}/g]$ is the typical
age-dependent scale of heat fluxes. Notice that, for fixed $x$ and $y$, $q$
grows like $t_w^{d/2-1}$. Therefore,
using the expression of $S$ given above, the extra term
$(NVg/4)[S^2(t)-S^2(t_w)]\propto (t_w^{-1}-t^{-1})\propto t_w^{-1}(1-x^{-1})$ is negligible
with respect to $q$, as anticipated.
The quantity $q$ is plotted in Fig.
\ref{fig-1} against $y$ for two different values of $x$. For any $x$
there is a negative minimum $-\beta_0^{-1} \equiv q(x,0)$ at the origin.
Since $q$ is the average heat exchanged by  the single modes, this
means that the components around $k=0$
cool as the time goes on and that
the cooling increases with  the time difference, as intuitively
expected. However, the shape of the curves shows that the rate of
cooling decreases as $y$ increases, with the unexpected and quite
interesting feature of the development of a positive peak
$\beta_M^{-1}\equiv q(x,y_M)$, which is more pronounced for the larger
time differences. This implies that the modes under the positive
peak warm up as time goes on. Since the thermal bath is at zero
temperature, this extra heat can only originate in the heat
redistribution due to the coupling among the modes. In fact, it
should be kept in mind that the linearization of the equations of
motion is only formal, the nonlinearity having been preserved
through the mean field term $S(t)$. For yet larger values of $y$ the
curves become flat about zero, indicating that the large $k$ modes
are equilibrated.

In order to see what are the implications of the above features on
the properties of $P(Q;t,t_w)$, let us compute the integral in
(\ref{p_di_q_con_modo_zero}).  Recalling that $b=0$, for $h$ in Eq.
(\ref{eqh}) to be defined, it must be $z\neq q(k,t,t_w)^{-1}$
$\forall k,t,t_w$. With the form (\ref{eqqsing}) of $q$ (see Fig.
\ref{fig-1}) this translates into $-\beta_0 < z_0 < \beta_M$. The
analyticity domain of $h$ is shown in the inset of Fig. \ref{fig-1}.
Notice that, in this far from equilibrium situation, $\beta _0$ and
$\beta _M$ play a role analogous to that of the inverse bath
temperature $\beta $ in equilibrium. We now compute the integral in
Eq. (\ref{p_di_q_con_modo_zero}) by steepest descent. The saddle
point $z^*$, if it exists, must satisfy the above constraint and the
saddle point equation ${\cal Q}=G(z;t,t_w)\vert_{z=z^*(t,t_w)}$, with
\begin{equation}
G(z;t,t_w)=\frac{1}{2}\int_{\Lambda}
\frac{d^dk}{(2\pi)^d}\frac{q(k,t,t_w)}{1-zq(k,t,t_w)}-gU(t,t_w). \label{saddle_point}
\end{equation}

Restricting the analysis to $2<d<4$ and using Eq. (\ref{eqqsing}),
one finds that $G(z=-\beta _0)\equiv {\cal Q}_c<0$ is finite, while
$G$ approaches infinity as $z$ tends to $\beta _M$. Therefore now
the saddle point equation admits a solution only for ${\cal Q}>
{\cal Q}_c$, and the integration  path is shown in the inset of Fig.
\ref{fig-1}. Instead, for ${\cal Q}\le {\cal Q}_c$, exploiting the
analyticity of $h$ in the neighborhood of the branch point $z=-\beta
_0$, an analysis similar to the classical one of \cite{berlkac}
shows that the steepest descent route deforms into a cusp
%\cite{lungo}
 whose peak is sticked in $z=-\beta_0$ (see inset of
Fig. (\ref{fig-1})). With this saddle point structure, finally one
obtains
%\be
%2\pi P(Q;t,t_w)=\exp [V h({\cal Q};z_{steep};t,t_w)],
%\ee
\begin{equation}
2\pi P(Q;t,t_w)=\exp [V h({\cal Q};z_{steep};t,t_w)]\, , \qquad
z_{steep}\,=\,\left\{\,\begin{array}{ccl}
z^*({\cal Q};t,t_w) \quad \mbox{for} \quad  {\cal Q}>{\cal Q}_c\\
-\beta_0 \qquad \quad \,\,\,\,\,\, \mbox{for} \quad  {\cal Q}>{\cal Q}_c\\
\end{array}\right.
\end{equation}
%where $z_{steep}=z^*$ for ${\cal Q}>{\cal Q}_c$ and $z_{steep}=\beta _0$
%elsewhere.
  The heat large-deviation
function ${\cal L} ({\cal Q};t,t_w)=h({\cal Q};z_{steep};t,t_w)$ is
plotted in Fig. (\ref{fig-2}).
Due to the sticking of $z_{steep}$, it consists of two parts. For
${\cal Q}\le {\cal Q}_c$ it is linear. For ${\cal Q}>{\cal Q}_c$ it
grows to a maximum and then falls off (again linearly for large
${\cal Q}$ since $z^*\to \beta_M$). The two branches merge at ${\cal
Q}_c$ with a discontinuity in  $d^3{\cal L}/d{\cal Q}^3$.

A singular behaviour qualitatively similar to that
described above has been observed in the numerical simulations of
the quenching dynamics of a disordered model for glassy systems
\cite{crisrit}.
This may suggest a certain generality
of the phenomenon in aging system and a possible common origin.
Non-analytical large deviation functions have been also
found  in a stochastic dissipative model for a single particle
\cite{farago}, in simple non-equilibrium systems coupled with two
reservoirs \cite{Visco,schutz}, and in diffusive models in the continuum
limit
%with infinitely many particles
%driven out of equilibrium by the boundaries
 \cite{jonasing,kafri}, always in stationary conditions.
The singular behaviour in \cite{farago, Visco, schutz} has been
related to the occurring of rare and very large fluctuations in the
initial distribution. 
%This may have some analogy with the case of the
%quenched system considered here, where the fluctuations
%of the disordered initial state are promoted to large
%inhomogeneities by the coarsening dynamics.
%Interestingly, a very similar behavior is also observed in numerical
%simulations of a random energy model for glasses \cite{crisrit}. In
%a different context, a singular large  deviation functional has been
%found in a driven stochastic particle model \cite{jonasing}. The
%possible existence of general conditions under which singular
%behavior occurs is an interesting matter for future research.

In the non-equilibrium setting considered here, 
the singular behavior of the distribution is related to the
tying of the saddle point solution to the analyticity edge. This
mechanism is mathematically similar to the one occurring in the
equilibrium phase-transition. In that context, the zero wave vector
fluctuations develop a macroscopic variance \cite{baxter,condvs}
through a mechanism reminiscent of the Bose-Einstein condensation. A
similar phenomenon is dynamically produced here in the realm of
fluctuating quantities: when a large amount $V {\cal Q} < Q_c \equiv
V {\cal Q}_c$ of heat is released, a macroscopic fraction $ Q - Q_c$
is provided by the $k=0$ mode. This is a novel condensation
mechanism for non-equilibrium fluctuations.

% A large fluctuation of the  heat released by
%the system can only occur if the $k=0$ mode
% (related to large inhomogeneity)
%can provide energy for it.
%This phenomenon in our model depends on
%the phase space structure and the value of the critical
%dimensionality has a role in it. An interesting perspective for
%future research is to investigate about the relations between the
%phenomenon of fluctuations condensation and that of condensation in
%equilibrium phase transitions.}

%We also ment ion that in the different context of particle models
%with infinitely degrees of freedom, driven out of equilibrium by the
%boundaries, large deviation singularities  have  been found in
%\cite{kafri} and in \cite{jonasing}, in presence of  strong bulk
%external  driving. In the last case the singularity  has been
%interpreted has a dynamical phase transition.

The large deviation function exhibits remarkable symmetry properties
in the limit $t_w\to \infty$ with ${\cal Q}$ fixed.
It can be shown
that in this limit the first term (i.e. $-z{\cal Q}$) in Eq.
(\ref{eqh}) is dominant, implying that the above limit
amounts to test the behavior of the tails of the heat
probability distribution.
In this regime one finds an expression where only
$\tilde {\cal Q}={\cal Q}/q_{typ}$ and $x=t/t_w$ appear \be {\cal
L}({\cal Q};t,t_w)= {\cal L}\,(\tilde {\cal Q};x), \label{scal} \ee
 with
${\cal L}(\tilde {\cal Q};x)=q_{typ}\beta _0 \tilde {\cal Q}$ or
${\cal L}(\tilde {\cal Q};x)=q_{typ}\beta _M \tilde {\cal Q}$ for
$\tilde {\cal Q}<0$ or $\tilde {\cal Q}>0$, respectively.
This shows
that in this process the same  scaling symmetry, which holds for
average quantities, underlies also the behavior of fluctuations. As
a consequence, ${\cal L}({\cal Q};t,t_w)$ takes the simple form of
Eq. (\ref{scal}) when its arguments are measured in units of their
reference value at the current age $t_w$ of the system. Moreover,
using the expression of ${\cal L}(\tilde {\cal Q};x)$ one finds the
asymmetry function
\be
{\cal L}({\cal Q};t,t_w)-{\cal L}(-{\cal
Q};t,t_w) =-(\beta_M-\beta_0){\cal Q}.
\label{fr1}
\ee
Plugging this result into Eq. (\ref{largedev}) one recovers a
relation formally identical to Eq. (\ref{fr}).
Notice however that the physical context is quite different:
Eq. (\ref{fr}) holds for $t-t_w$ large, while the validity of
Eq. (\ref{largedev}) requires $V$ large.
Apart from this difference, Eq. (\ref{fr1}) shows
that, by virtue of a scaling symmetry, a fluctuation relation like
(\ref{fr}) may be obeyed also in systems that are not at
stationarity, but are slowly relaxing and aging. To the best of our
knowledge, this is the first analytical result showing this in a
classical model of statistical mechanics with a non-trivial
equilibrium phase diagram. The quantities $\beta _0,\beta _M$
represent the origin of the cuts of $h$, and in close analogy to the
equilibrium case can be regarded as self-generated internal
temperatures. According to Eq. (\ref{fr1}) these regulate large heat
fluxes. Recalling that $-\beta _0=q^{-1}(k=0)$ and $\beta
_M=q^{-1}(k=k_M)$, such {\it temperatures} can be naturally
associated to the ordering modes releasing energy at $\vec k=0$, and
to those absorbing heat at a finite wave vector $\vec
k_M=y_Mt_w^{-1/2}$ (see Fig. \ref{fig-1}). Notice that, for $t_w\to
\infty$ and fixed $x$, $\beta _0 $ and $\beta _M$ decrease to zero
as $q_{typ}^{-1}=t_w^{1-d/2}$. Interestingly enough, this is the
same behavior observed for the so called {\it effective} temperature
$\beta _{eff}$, defined in terms of the ratio between the response
and the correlation functions, in the present model \cite{noi2002}.
However, the relation between the quantities $\beta _0$, $\beta _M$
entering $P(Q;t,t_w)$ and $\beta _{eff}$ remains to be fully
clarified.

Finally, we briefly discuss the modifications introduced to the
present picture by a quench to a finite temperature. It has been
shown \cite{noi2002} that, in this case, the order parameter can be
split into two statistically independent fields
$\boldsymbol{\varphi}=\boldsymbol{\sigma}+\boldsymbol{\psi}$, where
$\boldsymbol{\sigma}$ and $\boldsymbol{\psi}$ are, respectively, an
ordering and a thermal fluctuation component.  In the scalar case
($N=1$), these two terms are associated to the slow aging process
caused by the displacement of interfaces and to the fast spin
fluctuations with
%an
 equilibrium character inside the bulk of the
domains. This additive property amounts to the splitting
$C=C^{(\sigma)}+C^{(\psi)}$ of the correlation entering our
calculations. In \cite{noi2002} it is shown that $C^{(\psi)}$ is the
equilibrium correlation at the quench temperature $T$, while
$C^{(\sigma)}$ behaves as in a quench to $T=0$, apart from some
trivial non-universal constants. Then, in place of
Eq.(\ref{p_di_q_con_modo_zero}), one arrives at
\begin{equation}
 P(Q;t,t_w)= \int_{-i\infty+z_0}^{+i\infty+z_0} \frac{dz}{2\pi i}\,e^{V \left
[h^{(\sigma)} +h^{(\psi)}+ h^{(\sigma \psi)}\right ]},
\label{p_di_q_con_T}
\end{equation}
where $h^{(\sigma)}$ and $h^{(\psi)}$ are given by Eq. (\ref{eqh})
by setting the correlator $C=C^{(\sigma)}$ or $C=C^{(\psi)}$
respectively, and  $h^{(\sigma \psi)}$ is a function containing
cross products $C^{(\sigma)}C^{(\psi)}$. In the limit $t_w\to
\infty$ with fixed $x$, or alternatively with $t-t_w$ fixed, it is
possible to show that the cross-term $h^{(\sigma \psi)}$ can be
neglected. Hence  the heat probability   results as the convolution
\be
P=P^{(\psi)}\,*\,P^{(\sigma)}= (2\pi i)^{-1}\int
_{-i\infty+z_0}^{i\infty+z_0} dz\,e^{V \left [h^{(\psi)} +
h^{(\sigma)}\right ]}
\ee
of the fast and slow degree distributions.
$P^{(\sigma)}$  has the properties discussed insofar for
%regarding
the quench to $T=0$  while $P^{(\psi)}$
 is the equilibrium distribution  at the temperature $T$.
Notice that, in the regime $t_w\to \infty$ with $t-t_w$ fixed,
$h^{(\sigma)}$ is negligible and one remains with the equilibrium
distribution alone $P=P^{(\psi)}$.
 With these
behaviors, it can be shown that the saddle point structure described
above is not changed, except for a shift of the branch points at
$\beta _0, \beta_M$
%{\color{red} ma e' vero che c'e' lo shift? Se i
%tagli a T=0 e T ne 0 si sovrappongono, prevalgono quelli a T=0 e non
%c'e' shift}.
Then a singularity in the large deviation function
occurs at a temperature dependent ${\cal Q}_c(T) < 0$. As it can be
seen in Fig. \ref{fig-2}, the convolution with the equilibrium part,
produces a broadening of ${\cal L}$
 particularly for large $T$ and/or small $x$.
This convolution structure,  shown here for the first time, is
expected to be  very general in aging systems where a wide
separation of time scales occurs, and also appropriate for other
fluctuating quantities, beside $Q$. We notice that an analogue
property is not expected in critical quenches at $T=T_c$ where the
additivity
$\boldsymbol{\varphi}=\boldsymbol{\sigma}+\boldsymbol{\psi}$ is not
obeyed: The composition of equilibrium and off-equilibrium
fluctuations in this case remains an interesting issue to be
clarified.

By summarizing,  we have computed the exact asymptotic probability
distribution of the heat exchanged by a quenched ferromagnet
described by the large-$N$ model. A rich scaling structure emerges
where  heat, released by the  small wave vector ordering modes,
flows to components with finite wave vectors. The heat large
deviation function shows a non-differentiable behavior with a
singular threshold $Q_c$ signalling the onset of fluctuations
condensation at zero wave vector. Heat currents are governed by a
fluctuation relation analogue to the one obeyed in stationary
systems in contact with two baths, but here with two self-generated
temperatures $\beta _0$, $\beta _M$. It is a
 challenge to establish to what extent  the  scenario above
outlined is generic and holds also for  systems with finite $N$.

\acknowledgements

We acknowledge fruitful discussions with Felix Ritort, Sergio
Ciliberto and Leticia F. Cugliandolo. GG acknowledges support by
PRIN 2009SKNEWA.

\end{document}